\documentclass[sigconf]{acmart}

\AtBeginDocument{%
  }

\copyrightyear{2024}
\acmYear{2024}
\acmDOI{10.1145/3613904.3642132}
\acmISBN{979-8-4007-0330-0/24/05}
\setcopyright{acmlicensed}\acmConference[CHI '24]{Proceedings of the CHI Conference on Human Factors in Computing Systems}{May 11--16, 2024}{Honolulu, HI, USA}

\acmBooktitle{Proceedings of the CHI Conference on Human Factors in Computing Systems (CHI '24), May 11--16, 2024, Honolulu, HI, USA}

\acmPrice{15.00}

\usepackage{comment}

\usepackage{tabularx,booktabs}
\newcolumntype{C}{>{\centering\arraybackslash}X} 
\usepackage{lipsum}

\usepackage{url} 
\usepackage{makecell}
\usepackage{siunitx}

\usepackage{algorithm,algcompatible,amsmath}

\usepackage{tabu}

\definecolor{TODOcolor}{HTML}{fa9fb5}
\definecolor{FANLEIcolor}{HTML}{feb24c}

\definecolor{inputvariable}{HTML}{1b9e77}
\definecolor{outputvariable}{HTML}{7570b3}
\newcommand{\inputvariable}[1]{\textbf{\textcolor{inputvariable}{#1}}}
\newcommand{\outputvariable}[1]{\textbf{\textcolor{outputvariable}{#1}}}

\begin{document}

\title[Understanding Reader Takeaways in Thematic Maps]{Understanding Reader Takeaways in Thematic Maps Under Varying Text, Detail, and Spatial Autocorrelation}

\author{Arlen~Fan}
\authornote{Both authors contributed equally to this research.}
\email{afan5@asu.edu}
\author{Fan~Lei}
\authornotemark[1]
\email{flei5@asu.edu}
\author{Michelle~V.~Mancenido}
\email{mvmancenido@asu.edu}
\author{Ross~Maciejewski}
\email{rmacieje@asu.edu}
\affiliation{%
  \institution{Arizona State University}
  \streetaddress{P.O. Box 1212}
  \city{Tempe}
  \state{AZ}
  \country{USA}
  \postcode{}
}

\author{Alan M. MacEachren}
\email{maceachren@psu.edu}
\affiliation{%
  \institution{Pennsylvania State University}
  \streetaddress{Circle}
  \city{State College}
  \country{USA}}

\renewcommand{\shortauthors}{Fan et al.}


\begin{abstract}
Maps are crucial in conveying geospatial data in diverse contexts such as news and scientific reports. This research, utilizing thematic maps, probes deeper into the underexplored intersection of text framing and map types in influencing map interpretation. In this work, we conducted experiments to evaluate how textual detail and semantic content variations affect the quality of insights derived from map examination. We also explored the influence of explanatory annotations across different map types (e.g., choropleth, hexbin, isarithmic), base map details, and changing levels of spatial autocorrelation in the data. From two online experiments with $N=103$ participants, we found that annotations, their specific attributes, and map type used to present the data significantly shape the quality of takeaways. Notably, we found that the effectiveness of annotations hinges on their contextual integration. These findings offer valuable guidance to the visualization community for crafting impactful thematic geospatial representations.
\end{abstract}

\ccsdesc[500]{Human-centered computing~Empirical studies in visualization}
\ccsdesc[500]{Human-centered computing~Information visualization}

\keywords{Visualization, text, annotation, design, maps}

\received{20 February 2007}
\received[revised]{12 March 2009}
\received[accepted]{5 June 2009}

\maketitle

\section{Introduction}
\label{sec:introduction}

Text elements, such as titles, captions, and supplementary information, have been shown to improve accessibility~\cite{borkin2015beyond}, comprehension ~\cite{chen2022does,10297564, pinheiro2022charttext} and the speed of information conveyance\cite{ottley2019curious} in various domains of application (e.g., education, news reporting, online retail). However, the majority of the research studies on the addition of text elements have been focused on simple data visualizations, such as line charts~\cite{kim2021towards, stokes2022striking}. Design recommendations for augmenting text elements in thematic maps, as representations of complex, geospatial data, are lacking in the literature.

Thematic maps, which use symbols, colors, and patterns to portray statistical data, are widely used in domains that rely on geospatial information (such as urban planning~\cite{yeh1999urban}, public health~\cite{ricketts2003geographic}, environmental science~\cite{goodchild2003geographic}, etc.) often revealing interesting trends and patterns across spatial dimensions. Maps, in general, have also become popular in digital platforms. In 2018, it was estimated that maps comprise 30\% of all D3.js visualizations found on the internet~\cite{battle2018beagle}. Because geospatial visualizations are inherently complex, designing effective and engaging maps for digital consumption poses unique challenges in comparison to other popular visualization techniques (\textit{e.g.} line, bar, pie charts).

The objective of this work is to provide \emph{design guidelines} for integrating text information into thematic maps to optimize the quality of reader takeaways. To accomplish this objective, we examine the effects of text and map-related elements on reader takeaways and comprehension by running two experiments in the crowdsourcing platform Prolific.co. In contrast with previous work~\cite{kim2021towards,stokes2022striking}, we used a more comprehensive set of text (e.g., semantic level) and map-related attributes (e.g., map type) to reflect the unique complexities of summarizing and presenting geospatial data. The majority of these attributes and their interactions were purposely manipulated in the preparation of the maps being presented to the $N=103$ participants. Nine major hypotheses on the effect of text elements on the rated quality of participant takeaways were formulated and tested through regression-type analysis of categorical data~\cite{stroup2012generalized}. The results of the modeling procedure were analyzed and interpreted to determine which subset of variables had significant impacts on the quality of takeaways.



This study provides the following major contributions:
\begin{itemize}
    \item It is demonstrated, via empirical data, that the effect of thematic map type on the granularity and semantic level of reader takeaways is dependent on the level of spatial detail and the principled design of text elements;   
    \item It is shown how changes in the geographic detail of thematic maps, in combination with changes in the design of related text elements, affect the granularity and semantic level of reader takeaways;
    
    \item It is demonstrated through empirical evidence that a reader's behavioral inclination to consult either the visual or text elements of a thematic map can be influenced by manipulating these elements;
    
    \item Finally, we propose practical design guidelines when integrating text elements into thematic maps depending on the desired objectives on reader takeaways.
\end{itemize}

\section{Related Work}
\label{section:related_work}

This work is grounded on theoretical foundations from visualization and cartography. This section provides a brief overview of relevant prior work on the topic.

\subsection{Human Interaction with Text in Visualizations}
\label{subsection:human_factors}

The role of text elements in visualization remains a polarizing topic, with some work concluding that visual elements are more influential than text (\textit{e.g.}~\cite{kim2021towards,o2018testing}) while other work suggests the opposite (\textit{e.g.}~\cite{borkin2013makes,kong2018frames}). For example, Kim et al.~\cite{kim2021towards} showed that when both chart and caption emphasized a high-prominence feature, it was predominantly considered the main takeaway by readers \textit{i.e.}, readers were found less likely to use information from the text to form their takeaways. Similarly, O'Brien and Lauer~\cite{o2018testing} found that deceptive techniques (such as truncated or inverted axes) caused readers to misinterpret information even when paired with accurate explanatory text. These works suggest that a reader's attention is more naturally influenced by the visualization rather than textual attributes. 

In contrast, several studies support the view that text annotations can have a strong influence on a reader's focus, inferred takeaways, and preferences. Borkin et al.'s eye-tracking study~\cite{borkin2015beyond} found that titles and text are key elements in a visualization that help with recall, with readers fixating on text even if it did not appear in the title. Kong et al.~\cite{kong2018frames} found that the slant of the title in a visualization influenced the perceived main idea, with readers arriving at divergent takeaways from the same visualization. Ottley et al.~\cite{ottley2019curious}, also through an eye-tracking study, found that visualizations \emph{primarily} facilitated the identification of main topics. Once the main topics were identified, however, it was found that readers extracted information from the text annotations. Additionally, participants with a preference for visual information over text information still preferred charts with a higher number of annotations. Finally, Hearst and Tory~\cite{hearst2019would} examined individual preferences for visualizations when interacting with chatbots, finding that more than half of surveyed participants preferred seeing text annotations without visualizations.

A recent work closest in spirit to our study is Stokes et al.~\cite{stokes2022striking}, which examined the interplay between text and visual elements in line charts. Their work captured reader preferences and takeaways when using these two modalities and distilled major findings into design guidelines for integrating text annotations into line charts. However, this study only examined annotations in univariate line charts and used synthetic datasets with abrupt peaks and trends. In contrast, our study expands on Stokes et al.’s initial findings by addressing several limitations: (1) we explore a more complex visualization technique (maps) and its interplay with text annotations; (2) we use real-world datasets and apply accepted methods in thematic cartography to generate base maps. 
Through preliminary discussions with a cartography expert, it is hypothesized that the guidelines for line charts as recommended by Stokes et al. may not naturally extend to thematic maps, due to the nuanced complexities of geospatial datasets and related visualization techniques.
Although recent research by Stokes et al.~\cite{stokes2021give} supports using text-only options when presenting data information, we excluded this option due to the intricacy of maps and possible biases arising from converting geospatial data to text (this limitation is further discussed in Section~\ref{section:limitations}).

\subsection{Thematic Map Design}
A thematic map is a type of map that emphasizes the spatial distribution of a particular theme, topic, or attribute within a specific geographic area. Instead of showing traditional geographical features like rivers, mountains, and political boundaries, thematic maps use various visual elements such as colors, symbols, and patterns to represent data related to a specific subject (a theme)~\cite{dent1999cartography}. For example, a population density thematic map might use different shades of color to depict the concentration of people in different regions, with darker colors indicating higher population densities. Thematic maps are valuable tools for visualizing and analyzing spatial data for a wide range of applications, including urban planning, environmental studies, and social sciences~\cite{slocum2008thematic}.

\vspace{1.5mm} \noindent \textbf{Spatial Autocorrelation}: A critical visual task when using thematic maps is the identification of spatial clusters. Typically, \emph{color} changes have been used to provide visible clustering among regions that have similarities or differences. The measurement of \emph{spatial autocorrelation} with respect to an attribute of interest has been the focus of previous work on clustering map components. Some popular metrics proposed include: Join count statistic~\cite{cliff1970spatial} developed for binary variables based on the probability that a unit area belongs to the same class as its adjacent areas; Moran's I~\cite{moran1950notes} which considers pairwise products of deviations; and Geary's C~\cite{geary1954contiguity} which uses the sum of squared distances. In this study, Moran's I was selected due to its emphasis on the global detection of similarities among regions in contrast with Geary's C, which is more appropriate for detecting spatial heterogeneity.

\begin{figure*}[t]
	\centering	
	\includegraphics[width=\linewidth]{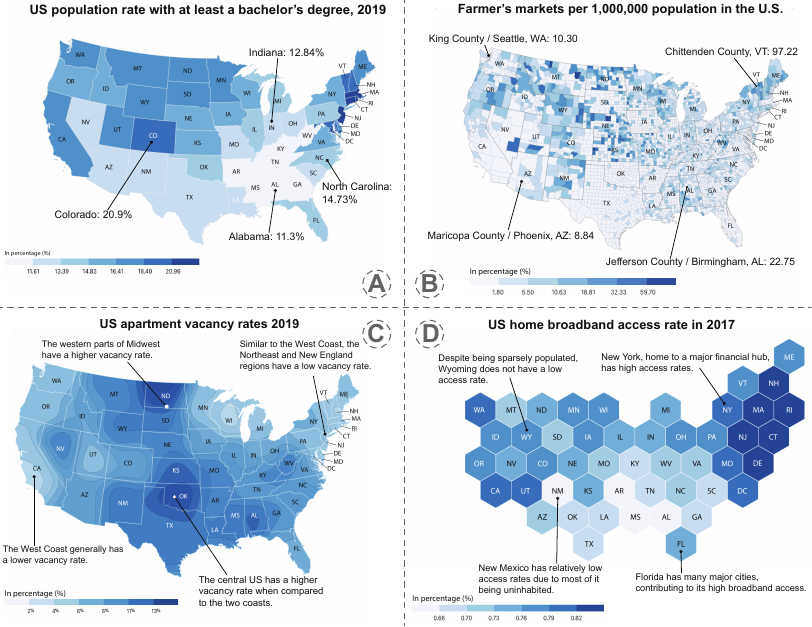}
	\caption{Different types of maps (which are bolded in this caption) with varying semantic levels of text as described by Lundgard and Satyanaryan~\cite{lundgard2021accessible} for classifying text in visualizations. A U.S. State-level \textbf{choropleth} map (A) and a county-level \textbf{choropleth} map (B) using descriptive statistics (Semantic Level 2 or perceiver-independent) text annotations. (C) An \textbf{isarithmic} map with text annotations describing complex and overall trends (Semantic Level 3 or perceiver-dependent). (D) A hexbin map, which is a \textbf{cartogram} variant, containing text annotations with external or background information (Semantic Level 4 or perceiver-dependent).}
	\label{fig:teaser}
    \Description{(4 Maps) This figure shows four different types of map stimuli with varying semantic levels of text. A shows a U.S. state-level choropleth map while B shows a U.S. county-level choropleth map. The text annotations in A and B contain the statistical information of some locations. C is an isarithmic map with text annotations describing regional trends. D shows a hexbin map containing external text information.}
\end{figure*}

\vspace{1.5mm} \noindent 
\textbf{Classification Scheme}: After identifying and calculating appropriate spatial autocorrelation metrics, the analysis naturally segues into choosing a classification scheme, a key aspect of thematic map design. While spatial autocorrelation measures the extent of spatial association among individual observations, classification schemes provide a set of rules for assigning each map region to a class. Selection of the classification scheme plays a major role in the appearance of spatial clusters, which could reveal key trends or patterns that would otherwise not be readily apparent~\cite{10273434}. For univariate data, the simplest class selection methods rely on quantiles and standard deviations~\cite{longley2005geographic}. More complex methods include Jenks’ natural breaks~\cite{jenks1971error}, which maximizes interclass variance and minimizes intraclass variance; Cromley’s~\cite{cromley1996comparison} minimum boundary error, which seeks to maximize spatial similarity among adjacent areas in the same class; and Jenks Caspall~\cite{jenks1971error}, which minimizes the deviation around class means using a heuristic process. Each method results in a different map appearance, which underlines the importance of the judicious selection of a classification scheme based on the statistical properties of the dataset. After careful consideration of the attributes of the dataset used in this study, we used the 7-class Fisher-Jenks method as our primary classification scheme for all generated maps.

\vspace{1.5mm} \noindent \textbf{Text in Maps}: In addition to graphical components, it is essential to highlight the distinct roles and importance of textual elements in maps, a notion supported by previous research \cite{fairbairn1993nature}. The effectiveness of textual elements largely hinges on their presentation. It is important to differentiate between text annotations or narratives, which supply pivotal additional content and explanations, and labels, which identify and delineate the elements they are attached. Traditionally, cartographic research has primarily centered on the study of text labels, focusing on enhancing their legibility and facilitating the correct association with the respective features they denote; a guidance substantiated in the detailed categorization by Dent et al.~\cite{dent1972visual}.

\vspace{1.5mm} \noindent 
\textbf{Data Storytelling and Data Journalism}: In recent years, accessible and interactive representations of geographic information called \emph{story maps} have also risen in popularity~\cite{roth2021cartographic}. Story maps integrate maps and text, organized in the form of focused narratives. They have been used as an engaging method for showing compelling evidence of the rise in global sea levels~\cite{song2022visual} and the spread of COVID-19~\cite{prestby2022design}. In the context of data journalism, Song et al.~\cite{song2022visual} studied whether themes (US presidential campaign donations, US coastal sea-level rise), genres (longform infographic, dynamic slideshow), or tropes (color highlighting, leader lines), would influence reader retention or comprehension. They found that the story theme had no influence, while participants performed better using long-form infographics and leader lines. Individual audience differences by expertise and prior beliefs also impacted participant response. Our work continues to build on these findings by closely examining the interplay between visual and textual elements in the comprehension process from a data visualization and cartography standpoint. Our examination will highlight the importance of proper text annotations in conveying clear and concise information and in balancing visual and textual elements for an optimal reader experience.

\section{Study Design}
\label{section:study_design}

In this section, we first define the manipulated variables in the study (Section~\ref{subsection:map_design_factors}). Then we state our research questions and hypotheses (Section~\ref{subsection:research_questions}). Finally, we explain our survey in detail (Sections~\ref{subsection:participants},~\ref{subsection:stimuli}, and~\ref{subsection:survey_measures})

\subsection{Map Design Factors}
\label{subsection:map_design_factors}

We considered best practices in map design and analyzed existing gaps to identify map design factors for evaluation. To inform the map design factors tested in this study, we consulted Slocum et al.'s~\cite{slocum2008thematic} textbook on thematic cartography and selected elements that could yield measurable variations in reader takeaways. Additionally, we drew insights from various other sources, including Stokes et al.'s~\cite{stokes2022striking} work to further inform the design of the study. Ultimately, we identified these five elements: Map Type, Map Detail, Semantic Level, Spatial Autocorrelation, and Text-Map Detail Alignment.

\vspace{1.5mm} \noindent \textbf{\raisebox{0pt}[0pt][0pt]{\raisebox{-0.5ex}{\includegraphics[height=2.5ex]{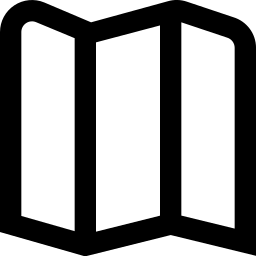}}} Map Type}: Map types, as a fundamental design factor, differ with respect to their purpose and distinct representations~~\cite{muehlenhaus2013web}. In this study, we limit our scope to common map types used for quantitative thematic mapping~\cite{dent1999cartography}.

Slocum et al.’s textbook [46] on thematic cartography categorizes three primary map types suitable for visualizing a univariate measure in a geographic context, each accompanied by relevant examples:

\begin{itemize}
    \item  \textbf{Choropleth Maps}: The most popular type of map for displaying statistical data for enumeration areas using an ordered color palette. A prototypical use of a choropleth map is to visualize population density or other demographic data derived for census enumeration units such as states or counties. An example  representing enumerated data is shown in Fig.~\ref{fig:teaser}A, B.
    \item  \textbf{Isarithmic Maps}: The second most popular map type, isarithmic maps display continuous data by connecting points or places of equal value with contoured lines. These lines divide the spatial surface into different areas and highlight where data levels change. Isarithmic maps can depict physically continuous surfaces (e.g., elevation or air temperature) or statistically continuous surfaces generated from enumerated data for areas (e.g., population density or average income). The latter use the same input data as choropleth maps and are the focus here. An example is shown in Fig.~\ref{fig:teaser}C.
    \item \textbf{Cartograms}: Similar to choropleth maps but the spatial geometry is distorted for visual effect. Cartograms are used in media and news, especially for political campaign analysis and sales data visualization. In this study, we opted for a specific variant of cartogram known as the hexbin map, where equal-area hexagons represent states in the U.S. While there are other types of cartograms, such as area and shape-warping, our focus is solely on the hexbin type since, like both the choropleth and isarithmic maps included in the study, they use color shading to depict univariate data values. An example is shown in Fig.~\ref{fig:teaser}D.
\end{itemize}

\vspace{1.5mm} \noindent \textbf{\raisebox{0pt}[0pt][0pt]{\raisebox{-0.5ex}{\includegraphics[height=2.5ex]{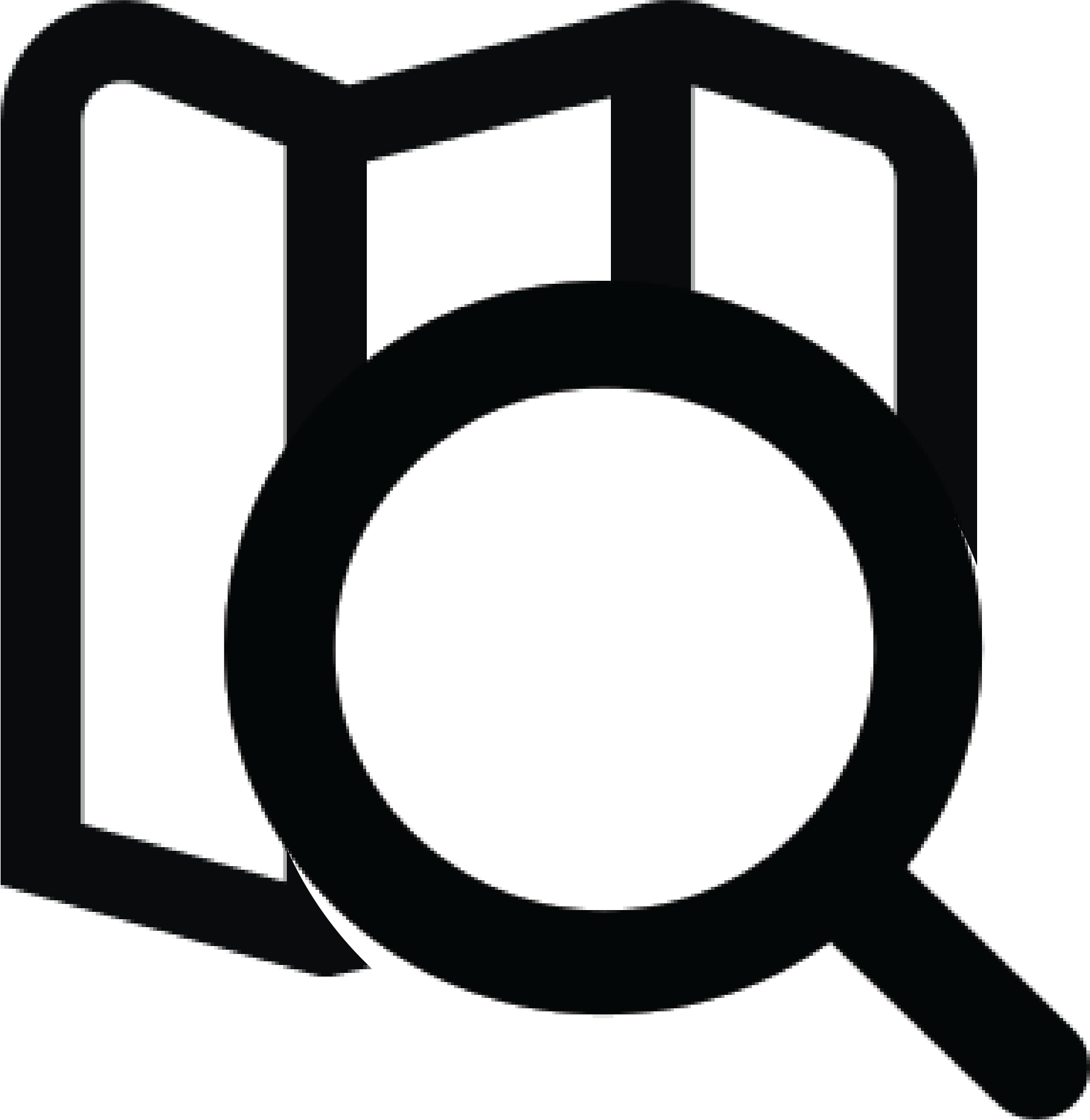}}} Map Detail}: A map's geographic detail has been known to influence its legibility and a reader's cognitive load for processing the presented information~\cite{aretz1991design}. While this factor~\cite{bunch2006cognitive} has been theoretically considered in past studies, its effect on reader takeaways has not been studied, specifically in the presence of changing text elements. In this study, we chose to use U.S. maps at the \emph{state} or \emph{county} level, with Alaska and Hawaii omitted. 

\begin{figure*}[t]
\centering	
\includegraphics[width=0.9\linewidth]{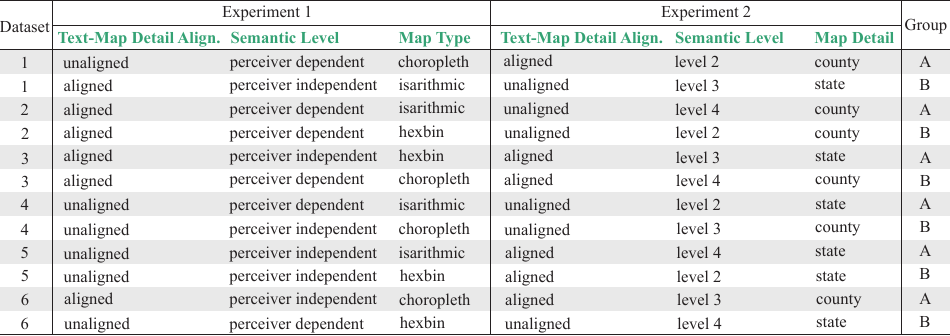}
\caption{Experimental design constructed using JMP\circledR software. The input variables—comprising text-map detail alignment, semantic level, map type, and map detail—were systematically varied to rigorously assess both individual and interaction effects. JMP\circledR was employed to optimize the design of experiments, ensuring comprehensive evaluation of all interaction effects. In Experiment 1, all of the maps are at the state level. In Experiment 2, all maps are choropleth maps, which means that the variable \textbf{\inputvariable{Map Type}} is constant.}
\label{tab:table-studydesign}
\Description{(Table) A table shows the input variables for two experiments as described in Section 3. The input variables are dataset, text-map detail align, semantic level, map type, map detail, and group.}
\end{figure*}

\vspace{1.5mm} \noindent \textbf{\raisebox{0pt}[0pt][0pt]{\raisebox{-0.5ex}{\includegraphics[height=2.5ex]{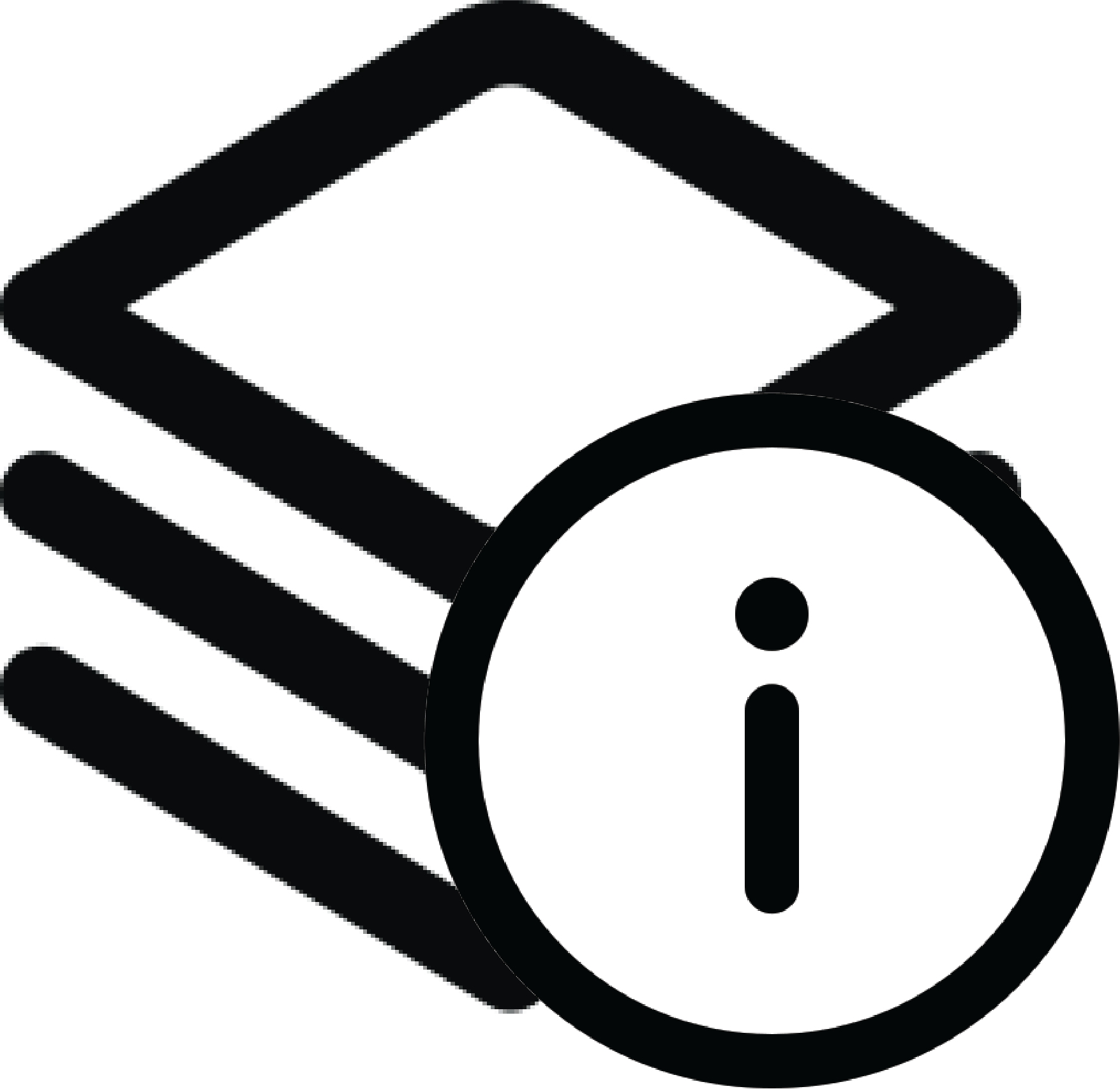}}} Semantic Level}: We apply Lundgard and Satyanaryan's framework~\cite{lundgard2021accessible}, which classifies accompanying text for visualizations into four levels. Originally intended to improve text accessibility for the visually impaired, Stokes et al.~\cite{stokes2022striking} utilized this framework to examine different aspects of a visualization's content.

\begin{itemize}
    \item \textbf{Semantic Level 1 (L1)}: Refers to elemental and encoded properties, such as chart type, encoding channels, title, and labels. L1 text is considered to be \textbf{perceiver-independent} i.e., perceptions or interpretations are not expected to change from one reader to the next.
    \item \textbf{Semantic Level 2 (L2)}: Covers descriptive statistics such as outliers, extrema, or point-wise comparisons. Similar to L1, text at this level is \textbf{perceiver-independent}.
    \item \textbf{Semantic Level 3 (L3)}: Describes perceptual and cognitive aspects, such as trends and patterns. Text at this level is considered to be \textbf{perceiver-dependent}, i.e., information inferred is contingent on a reader's interpretation and perception.
    \item \textbf{Semantic Level 4 (L4)}: Refers to external information, such as past and current events that supplement the topic. Similar to L3, these types of annotations are \textbf{perceiver-dependent}.
\end{itemize}

\noindent \textbf{\raisebox{0pt}[0pt][0pt]{\raisebox{-0.6ex}{\includegraphics[height=2.5ex]{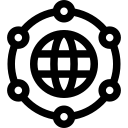}}} Spatial Autocorrelation}: Changes in spatial autocorrelation result in changes in map appearance, as well as information regarding associations among adjacent locations~\cite{cliff1981spatial,legendre1993spatial}. Examining spatial autocorrelation as a factor in map perception allows for the examination of how readers perceive spatial patterns and clusters, which is crucial for understanding spatial cognition. A pioneering study by Olson~\cite{olson1975autocorrelation} examined the relationship between people's perception of complexity and autocorrelation and clustering. The experiment found that highly clustered visualizations were generally rated as less complex, while dispersed visualizations were seen as more complex. Participants' open-ended feedback supported these findings. Later, Bunch and Lloyd~\cite{bunch2006cognitive} explored various types of cognitive load in the context of geographic information, distinguishing between intrinsic load (associated with inherent complexity), extraneous load (related to presentation format), and germane load (linked to processing and schema automation). Given these insights, we hypothesize that spatial autocorrelation influences reader understanding.

In this paper, we use the metric Moran's I~\cite{moran1950notes}, which calculates the degree of similarity between neighboring observations to measure a map's global spatial autocorrelation. Moran's I is sensitive to broad patterns and is ideal for identifying overarching trends in spatial datasets, making it a suitable measure of spatial autocorrelation for the purposes of our study.

\vspace{1.5mm} \noindent \textbf{\raisebox{0pt}[0pt][0pt]{\raisebox{-0.5ex}{\includegraphics[height=2.5ex]{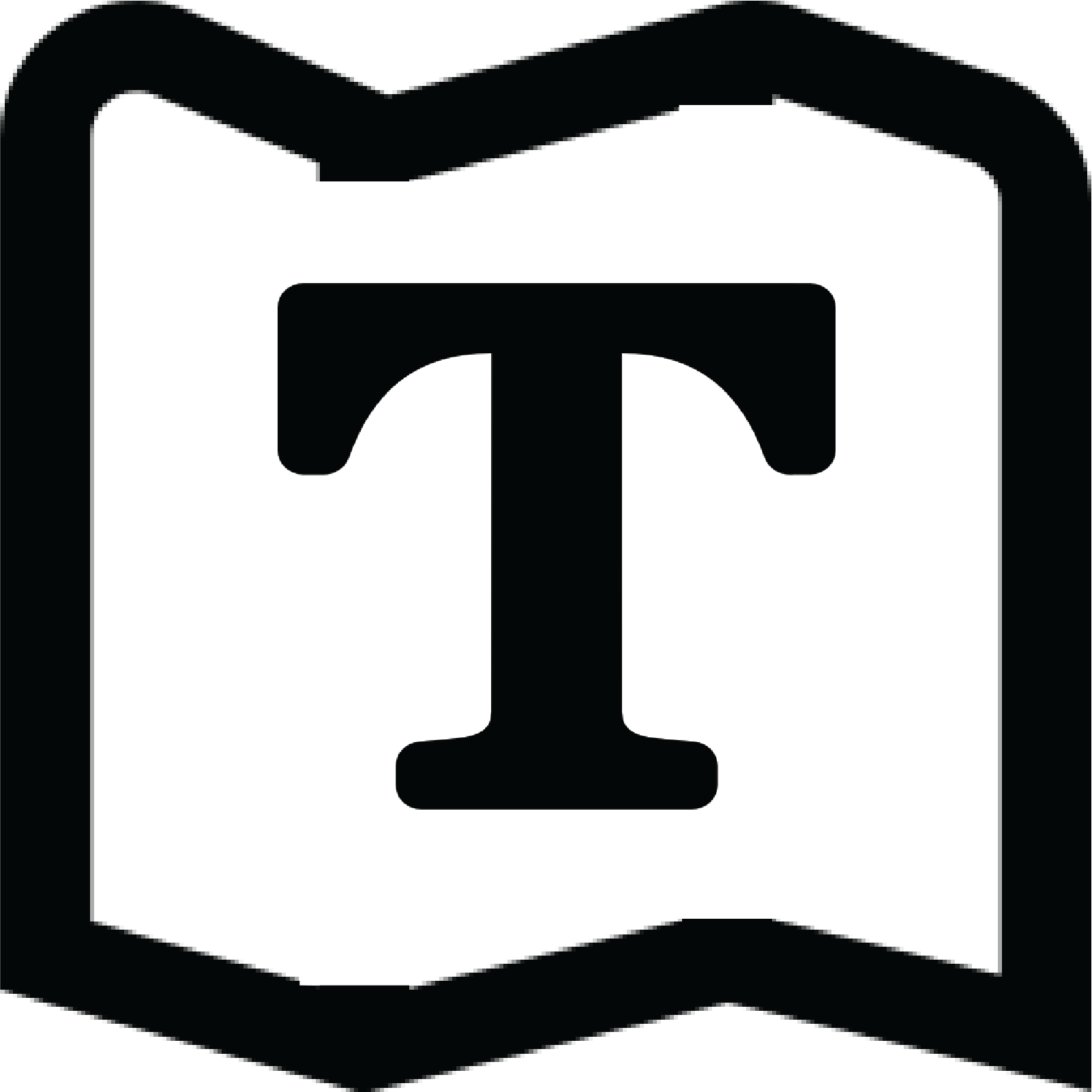}}} Text-Map Detail Alignment}: This variable focuses on the relationship between the granularity of text annotations and the corresponding map detail. When the level of detail in text annotations matches that of the map, such as county-level text on a county-level map or state-level text on a state-level map, it is considered aligned. Conversely, misalignment occurs when the granularity of text annotations does not correspond with the map detail. For instance, county-level text annotations on a state-level map represent a misalignment. Our interest in this variable stems from a hypothesis that the alignment (or misalignment) of these elements is not merely a design choice, but a factor that can lead to interesting and measurable differences in how readers process and understand thematic maps. We included this variable to primarily explore cognitive processing in map reading, as misaligned text and map detail may pose cognitive challenges, requiring readers to integrate separate pieces of information.

To effectively author annotations, we consulted  Fairbairn's~\cite{fairbairn1993nature} paper, which lists a comprehensive taxonomy of text symbols that can appear on maps. They can be (1) descriptive, which reflects features that are symbolized on the map; (2) analytical, which links the reader with feature attributes; and (3) positional, which describes or confirmations location. These text types have further subcategories. However, as our study only utilizes univariate thematic maps, many of these text types are not applicable. For instance, it would be unsuitable to include warning text or longitude and latitude markings on choropleth maps. Consequently, we will use descriptive (L4), determinative (L2), interpretative (L3-L4), and temporal/positional (L4) text in the study.

\begin{figure*}[t]
	\centering	 
	\includegraphics[width=\linewidth]{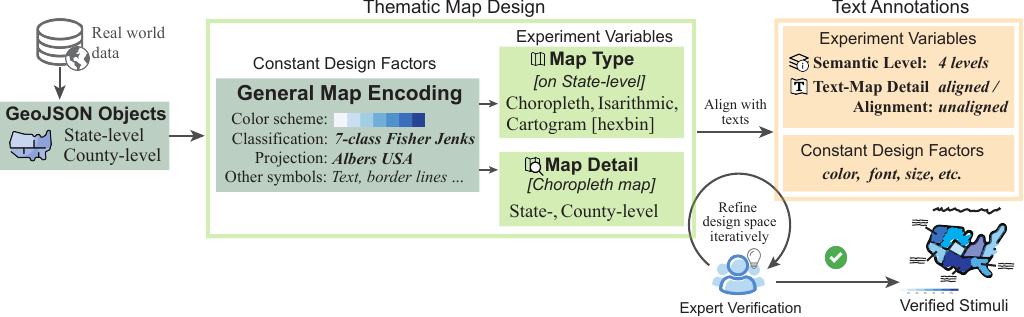}
	\caption{The stimuli construction pipeline. Data were mapped onto GeoJSON objects and thematic map design principles were applied to generate all stimuli. Map design factors were either varied or kept constant as required by our experimental design (Fig.~\ref{tab:table-studydesign}). Maps were validated for adherence to best practices in thematic map design. Finally, annotations were added with differing semantic levels and reviewed a final time to produce verified stimuli.}
	\label{fig:stimuli_pipeline}
 \Description{(Flowchart) This figure shows the stimuli construction pipeline as described in Section 3.4. We first built the general map encoding space by working with a cartographer. Then we generated different map stimuli by varying experiment input variables. After generating the base map, we added text annotations on the map by controlling the experiment input variables. At last, the domain expert verified the stimuli designs.}
\end{figure*}

\subsection{Research Questions}
\label{subsection:research_questions}

While past work~\cite{griffin2005does} has focused on exploring the effects of text and map design factors on a reader's comprehension, our work differs in that we considered the interdependence of these factors on novel measures of design effectiveness via the takeaways provided. First, the participants were asked to formulate a minimum of 1 to a maximum of 5 takeaways, where each takeaway is a set of sentences describing the most important information gleaned from the annotated map. Then we asked participants to rate the perceived influence of map and text in their process of obtaining information from an annotated map. The takeaways were then rated by the project team with respect to the level of geographic detail of information provided (county, state, or region) and semantic level. Thus, the following research questions and related hypotheses focused on the effects of the design variables on these three performance metrics: source, detail, and semantic level of takeaways. 

\vspace{1.5mm} \noindent  \textbf{Research Question 1 (RQ1):} 
\emph{How does the design of a map influence a reader’s primary source for information (map or text annotations)?}

\begin{itemize}
    \item \textbf{Hypothesis 1a (H1a):} \textbf{Map type influences a reader's reliance on text annotations.}
        \begin{itemize}
            \item \emph{Explanation:} Distorted maps such as cartograms might lead to increased reliance on textual information due to their unfamiliar graphical representations. The semantic level of text annotations could further modulate this effect.
        \end{itemize}

    \item \textbf{Hypothesis 1b (H1b):} \textbf{Increased geographic detail in maps leads to greater reliance on text annotations.}
        \begin{itemize}
            \item \emph{Explanation:} Finer geographic detail can increase cognitive load, potentially enhancing reliance on high-level semantic text annotations for clearer understanding.
        \end{itemize}

    \item \textbf{Hypothesis 1c (H1c):} \textbf{Higher spatial autocorrelation in maps might reduce reliance on text annotations.}
        \begin{itemize}
            \item \emph{Explanation:} Maps with high spatial autocorrelation tend to exhibit clearer patterns, possibly reducing the need for text annotations. The map type might alter this effect.
        \end{itemize}
\end{itemize}

\noindent \textbf{Research Question 2 (RQ2):} 
\emph{How does map design affect the granularity of takeaways considering the map type, geographic detail, and text content?}

\begin{itemize}
    \item \textbf{Hypothesis 2a (H2a):} \textbf{Map type affects the granularity of a reader's takeaways.}
        \begin{itemize}
            \item \emph{Explanation:} Different map types like isarithmic maps may draw attention to broader regions, influencing the level of detail in the takeaways.
        \end{itemize}

    \item \textbf{Hypothesis 2b (H2b):} \textbf{The granularity of takeaways varies with map detail.}
        \begin{itemize}
            \item \emph{Explanation:} The level of detail in both the map and text can influence the specificity of a reader’s interpretation, with consistency between the two reinforcing certain levels of detail.
        \end{itemize}

    \item \textbf{Hypothesis 2c (H2c):} \textbf{Higher spatial autocorrelation leads to takeaways with coarser detail.}
        \begin{itemize}
            \item \emph{Explanation:} Spatially autocorrelated data often forms clusters, steering readers towards broader, cluster-based interpretations.
        \end{itemize}
\end{itemize}

\noindent \textbf{Research Question 3 (RQ3):} 
\emph{How does map design impact the semantic level of takeaways in relation to geographic detail and text content?}

\begin{itemize}
    \item \textbf{Hypothesis 3a (H3a):} \textbf{Map type influences the semantic level of a reader’s takeaways.}
        \begin{itemize}
            \item \emph{Explanation:} Certain maps, like isarithmic ones, may encourage higher-level semantic interpretations due to their focus on broader regions.
        \end{itemize}

    \item \textbf{Hypothesis 3b (H3b):} \textbf{Coarser map details lead to higher-level semantic takeaways.}
        \begin{itemize}
            \item \emph{Explanation:} Describing phenomena across larger areas typically falls into higher semantic levels, with the strength of this effect influenced by the semantic level of text annotations.
        \end{itemize}

    \item \textbf{Hypothesis 3c (H3c):} \textbf{Spatial autocorrelation within a map dataset influences the semantic level of takeaways.}
        \begin{itemize}
            \item \emph{Explanation:} Higher spatial autocorrelation, particularly in maps like isarithmic ones, can lead to more complex and higher-level semantic interpretations.
        \end{itemize}
\end{itemize}

All hypotheses and analyses scripts are preregistered on OSF~\footnote{https://osf.io/3pzvj/?view\_only=86167a2f0ba041899f6edee0c73c6f68. This is an anonymous, read-only link}.

\subsection{Participants}
\label{subsection:participants}

Participants for our study were recruited via Prolific.co, with eligibility criteria including U.S. residency, desktop computer usage, and a minimum 95\% task acceptance rate on Prolific.co. The survey, designed for a 20-minute completion time, offered a \$4 compensation upon successful completion, equating to an hourly rate of \$12. We considered 103 participants in our results (65 male, 36 female, 2 non-binary, $M_{age} = 37.2, SD_{age} = 23.3$). The majority (61) had at least a 4-year degree.

\subsection{Stimuli}
\label{subsection:stimuli}

The experimental design, as depicted in Fig.~\ref{tab:table-studydesign}, was developed using the JMP® statistical analysis tool, where \textbf{Map Type}, \textbf{Map Detail}, \textbf{Spatial Autocorrelation}, \textbf{Semantic Level}, and \textbf{Text-Map Detail Alignment} were considered to construct an efficient experimental design. We began by assembling six real-world datasets at both the state and county levels.

\begin{figure}[t]
\centering	
\includegraphics[width=0.9\linewidth]{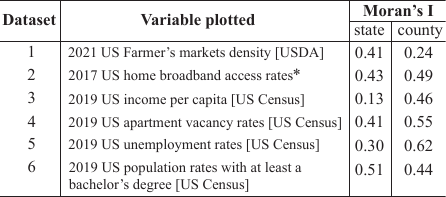}
\caption{The 6 datasets used in the study. The Moran's I value was calculated by using the choropleth map. *Dataset 2 is from Tolbert et al.~\cite{tolbert2015us}}
\label{table:table-dataset}
\Description{(Table) A table shows six datasets containing socioeconomic data of the United States. For each dataset, the table presents the global Moran's I values of the State-level and County-level spatial distribution.}
\end{figure}

Generating the stimuli followed a structured pipeline, visually represented in Fig.~\ref{fig:stimuli_pipeline}. The pipeline consisted of several key stages: encoding the assembled data into GeoJSON objects, applying thematic map design principles, and introducing variations or constants in map design factors as dictated by our experimental design (Fig.~\ref{tab:table-studydesign}). An important consideration was the spatial autocorrelation metric, Moran's I, which is determined by the dataset properties. We documented the values of Moran's I for both the state and county levels as indicators of spatial autocorrelation for each dataset (Fig.~\ref{table:table-dataset}).

\subsubsection{Thematic Map Design}
\label{subsubsection:thematic_map_design}

Thematic map design factors not measured in our experiment (\textit{e.g.} color, classification) must be controlled, as they have an effect on reader perception and introduce biases. Thus, we use the same color scheme and classification on all of our maps. We considered multiple classification division methods for our maps and evaluated their performance using Smith's~\cite{smith1986comparing} Goodness of Variance Fit (GVF). After weighing our options, a 7-class Fisher-Jenks~\cite{jenks1971error} was selected, as it consistently was in the top quartile of GVF scores. The legend is divided into 7 data ranges, and the class breaks were defined by the Fisher Jenks classification criteria~\cite{rey2017evaluation}, which minimizes deviation around class means using a heuristic process. We chose ColorBrewer's~\cite{harrower2003colorbrewer} 7-class sequential blue color scheme to depict the classes that the method defined.

After selecting the color scheme and classification method, we encoded symbols (\textit{e.g} border width, stroke, and typeface) onto the map. Following design guidelines from Dent et al.~\cite{dent1999cartography}, we added states standard two-letter abbreviations using  18pt Arial font and drew state borders with a 0.5pt width. For states that are too small to fit the text, solid black leader lines with 1pt width were used. All maps used the Albers USA projection~\cite{snyder1989album} except for the hexbins. After unifying the general map design space, we generated thematic maps by varying the \textbf{\inputvariable{Map Type}}, which took on one of three possible values: choropleth map, isarithmic map, or hexbin map. All maps were implemented in D3.js~\cite{bostock2011d3} and exported into the SVG format.

\vspace{1.5mm} \noindent  \textbf{\textit{Choropleth map:}} For each dataset, we generated state-level (Fig.~\ref{fig:teaser}A) or county-level choropleth maps (Fig.~\ref{fig:teaser}B) by varying the \textbf{\inputvariable{Map Detail}}. We grouped them into subsets (see Fig.~\ref{tab:table-studydesign}).

\vspace{1.5mm} \noindent  \textbf{\textit{Isarithmic map:}} Isarithmic maps, Fig.~\ref{fig:teaser}C, are used to illustrate the density of the observations on a map~\cite{dent1999cartography}. To convert discrete area data into continuous contour data, we use Kernel Density Estimation (KDE) on county-level choropleth maps. For each isarithmic map, we employed the fixed bandwidth KDE as described by Wang et al.~\cite{wang2014s}.

\vspace{1.5mm} \noindent  \textbf{\textit{Cartogram:}} Cartograms are widely used by media for storytelling. However, due to the absence of strict sorting criteria for grid map layouts used in data journalism, there are numerous varations~\cite{9827962}. In this case, we applied a popular grid layout practice from NPR's Danny Debelius~\cite{debelius_2015} which uses tessellated, equal-area hexagons (Fig.~\ref{fig:teaser}D) to represent states. As stated earlier, we use the hexbin variant exclusively in the study.

We recognized that certain combinations of map types and annotations might not be as effective, such as hexbin maps paired with annotations discussing aggregate patterns in regions like the western parts of the Midwest. Our methodology accounted for this, hypothesizing that different map types might have varying suitability for annotations at particular semantic levels.

\subsubsection{Annotation Content and Placement}
\label{subsubsection:annotation_content}

The \textbf{\inputvariable{Semantic Level}} refers to the framework outlined by Lundgard and Satyanaryan~\cite{lundgard2021accessible}. L1 text, such as title and legend, was present in all map stimuli. Annotations were manually authored in order to ensure that the text content was consistent with the semantic level described. For L2 annotations, descriptive statistics were used (See Fig.~\ref{fig:teaser}A).

\begin{figure}[b]
	\centering
	\includegraphics[width=\linewidth]{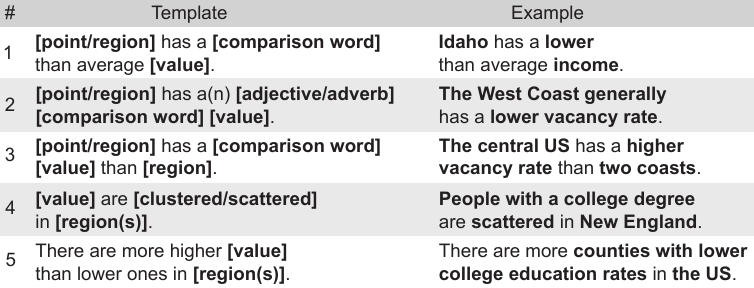}
	\caption{Templates used for L3 annotations.}
	\label{fig:l3_annotation_temp}
    \Description{(Table) This figure shows the text template and corresponding examples for L3 annotations as Section 3.4.2 illustrates.}
\end{figure}

\begin{figure*}[t]
	\centering	
	\includegraphics[width=\linewidth]{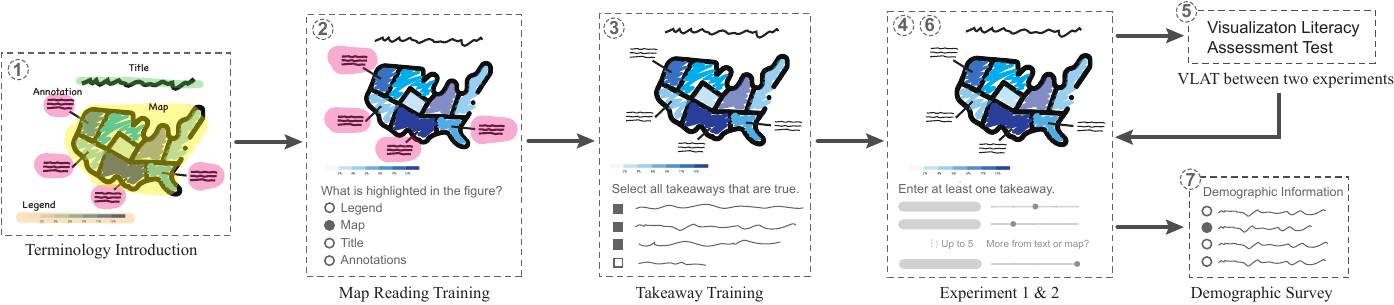}
	\caption{Participants workflow with seven sections. 1) terminology introduction 2) map reading training, 3) takeaway introduction and training, 4) the first experiment investigates how different map types interacted with varying textual elements, 5) VLAT, 6) the second experiment examines the interplay between choropleth maps' varying geographic detail and textual elements, and 7) demographics.}
	\label{fig:workflow}
    \Description{(Flowchart) This figure shows the participants' workflow when they attend the experiment. The workflow contains seven sections as described in Section 3.5.}
\end{figure*}

Aware of the potential influence of titles and framing effects on interpretation and takeaways, as discussed in Hullman and Diakopoulos's work on visualization rhetoric~\cite{hullman2011visualization}, we aimed to keep titles impartial, stating only the statistic being plotted without additional commentary or framing. This strategy was intended to minimize any bias. Additionally, our map legends were deliberately left unlabeled, as the titles themselves were designed to directly convey that information.

We used the text templates for L3 annotations as Fig.~\ref{fig:l3_annotation_temp} shows. Templates \#1 and \#2 focus on perceiver-dependent phenomena or subjective, imprecise descriptions of the data, whereas \#3 compares individual points to multiple points or entire regions. Templates \#4 and \#5 address complex trends and observations. To develop L4 annotations, we searched Wikipedia for relevant articles and chose information that offered suitable explanations for map areas. For perceiver-dependent annotations, we used either L3 or a mix of L3 and L4, with each map containing four annotations. 

We drew leader lines manually, as no practical implementation was identified in the existing literature. We considered creating annotations that coincided with visually salient features as demonstrated in Kim et al.'s study~\cite{kim2021towards}; however, our research lacks a corresponding crowdsourcing component. Still, we followed Dent et al.'s map construction guidelines~\cite{dent1999cartography} to place critical elements such as the title, legend, and map itself. Starting with a blank canvas, we placed each element and calculated the remaining workable map space, repeating this process until all elements were situated, resulting in a visually balanced map.

\subsection{Survey Measures}
\label{subsection:survey_measures}

The full survey pipeline is detailed in Fig.~\ref{fig:workflow}. Before the study starts, the participants are either assigned to Group A or B. The exact stimuli seen for each group are in Fig.~\ref{tab:table-studydesign}. The first section was an attention check and terminology introduction. Participants were shown a map with four basic components of thematic maps: legend, title, map, and annotations. After seeing this introduction page, in the training section, they were asked to identify two of the four components at random. Participants must answer both correctly to continue.

As part of our study, we focused on gathering data on specific output variables to understand how participants interact with and interpret the map stimuli. Our primary output variables are:

\vspace{1.5mm} \noindent \textbf{\outputvariable{Source}}: An ordinal measure of where the reader obtained the takeaway. This variable can assume five different values, representing the source of the reader's takeaway; a value of 1 denotes deriving insight solely from the text, while a value of 5 signifies solely relying on the map. Values between 1 and 5 indicate varying degrees of combined influence from both the text and the map.

\vspace{1.5mm} \noindent \textbf{\outputvariable{Takeaway Granularity}}: This variable captures the level of detail in the participants' takeaways, categorized into county, state, or region.

\vspace{1.5mm} \noindent \textbf{\outputvariable{Takeaway Semantic Level}}: Assessing the semantic depth of the takeaways, classified according to the framework outlined by Lundgard and Satyanaryan~\cite{lundgard2021accessible}.

The third section was a continuation of the training, primarily designed to inform the participant what the takeaways are and how to form them after reading a visualization. We defined a takeaway as ``a key fact, point, or idea to be remembered after viewing the map'', adapted from Stokes et al.’s~\cite{stokes2022striking} study. We provide a list of examples of takeaways and a brief explanation of how they were formed. We clarified that a reader can form a takeaway from reading the annotations about the map, reading the map only, or a combination of both. After this training, participants were asked two multiple-choice questions. There were four answer choices and three of them were correct. The incorrect answer was a direct contradiction to one of the annotations.

The fourth section of the study consisted of the stimuli in Experiment 1. Each participant saw six maps and was asked to write their takeaways for each map. They were allowed a maximum of five takeaways and a minimum of one takeaway per map. For each takeaway, participants reported how they formed their takeaways on a 5-point scale (1- all from text, 2- more text, 3- same, 4- more map, 5- all from the map), which constitutes the \textbf{\outputvariable{Source}} variable.

\begin{figure}[b]
	\centering
	\includegraphics[width=\linewidth]{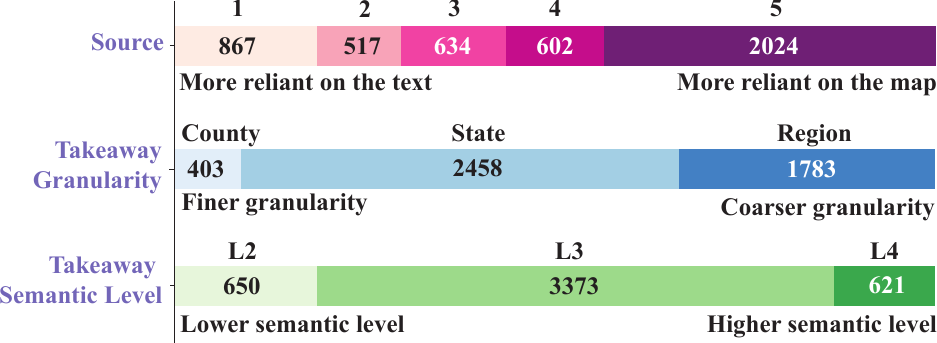}
	\caption{A statistical count of all of the coded takeaways. Horizontal bars stack to 4,644, the total number of takeaways analyzed in the study. \textbf{\outputvariable{Source}} took on 5 values on a 5-point scale (1- all from text, 2- more text, 3- same, 4- more map, 5- all from the map)}
	\label{tab:takeaway-counts}
    \Description{(Stacked Bar Charts) This figure shows the overview of all 4,644 takeaways we collected from the participants. These takeaways were coded by the takeaway granularity, the takeaway semantic level, and the source. The source variable can assume five different values, indicating the origin of the reader's takeaway; a value of 1 signifies that the takeaway stemmed exclusively from the text, while a value of 5 denotes that it was obtained entirely from the map.}
\end{figure}

\begin{figure*}[t]
\centering	
\includegraphics[width=\linewidth]{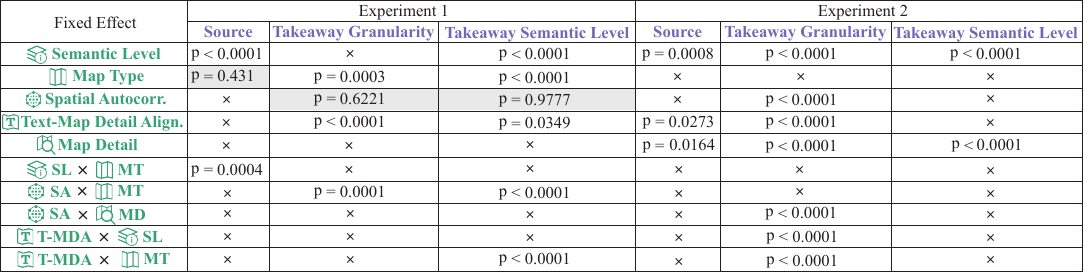}
\caption{Table of Fixed Effects. All independent variables are \inputvariable{bolded and colored green} and dependent variables are \outputvariable{bolded and colored purple}. \textbf{\inputvariable{Semantic Level}} was abbreviated to SL, \textbf{\inputvariable{Map Type}} to MT, \textbf{\inputvariable{Spatial Autocorrelation}} to SA, \textbf{\inputvariable{Text-Map Detail Align}} to T-MDA, and \textbf{\inputvariable{Map Detail}} to MD. Cells containing p-values indicate that the particular result was tested in our hypotheses. Grayed-out boxes indicate that the result is not significant at the $\alpha=0.05$ level. Crossed-out boxes indicate that the result is discarded due to non-significance or is not of interest.}
\label{tab:fixed-effects}
\Description{(Table) This figure is a table of Fixed Effects among the manipulated factors (texts in green) and response variables (texts in purple) as demonstrated in Section 4.1.}
\end{figure*}

The fifth section of the study was an abridged version of the Visual Literacy Assessment Test~\cite{lee2016vlat}, a validated way of assessing critical reading of various data visualizations such as line, pie, bar, stacked bar, etc. Since our study is primarily concerned with spatial data, we used questions pertaining to scatterplots, bubble charts, and choropleth maps, for a total of 17 questions. The purpose of this section was to reduce stimuli fatigue by adding variance to the tasks. It is also to properly test participants on their ability to critically read spatial data visualizations in lieu of relying on a self-reported metric. The sixth section of the study consisted of the stimuli in Experiment 2. Mechanics from Experiment 1 were repeated. Finally, the seventh and final section was a demographic survey.

\section{Results}
\label{section:results}

Two authors independently rated participant takeaways for the level of detail (county, state, region) and the semantic level. One hundred and three ($N=103$) participants provided a total of 4,644 takeaways, an average of 45.09 takeaways for each participant. There were 55 Group A participants, who provided a total of 2,534 takeaways total and 46 Group B participants, who provided 2,110 takeaways in total. The coders disagreed on the level of detail in 3.15\% of all responses and on the semantic level in 6.72\% of all responses. A third coder resolved all of the remaining conflicts except for 9. The remaining conflicts were discussed among the three coders. Fig.~\ref{tab:takeaway-counts} shows an overview of the coded takeaways.

Out of the total takeaways, 44 from 15 distinct participants were categorized as off-topic. These were further subdivided into three categories: 19 were classified as blanks, which were zero-character answers submitted for the sole purpose of advancing the survey questions; 4 were related to technical difficulties, such as issues with loading the map ("The image doesn't load and I'm not allowed to refresh the page..."); and 21 were statements that did not qualify as takeaways. These non-takeaways often involved participants merely repeating the title or did not meaningfully relate to the dataset, and they could not be easily classified under semantic levels L2, L3, or L4.

\subsection{Model Building}

Statistical analysis was conducted using the SAS$\circledR$ statistical software package. Due to the non-continuous nature of some of the variables, such as \textbf{\inputvariable{Map Detail}}, we used a flexible family of models called generalized linear mixed models (GLMM) for both estimation and tests of significance~\cite{stroup2012generalized}. The SAS$\circledR$ software's GLIMMIX package was used to estimate and test different GLMM's. We initially fitted models that had active main effects and two-factor interactions (2FI); then, we used $F$-tests to determine and eliminate effects that had no significant impact on the response (\outputvariable{Source}, \outputvariable{Takeaway Granularity}, \outputvariable{Takeaway Semantic Level}) via the $p$-value criterion. Finally, the final GLMM used for analysis only consisted of main effects and/or 2FI that had $p$-values less than the significance level of 0.05. We believe that these model forms, specifically with the addition of the 2FI, were parsimonious enough to be interpretable but complex enough to capture interesting joint effects of the variables of interest.

Post-hoc analysis of the difference in contrasts between groups (i.e., categories of the independent variables) yielded estimates of odds ratios. For binary responses, odds are the ratios of the probabilities of success vs. failure; in the case of ordered data, the odds are proportional odds which are equal odds (with respect to any two categories) of the outcome being a higher-level vs. a lower level category. Odds ratios, on the other hand, are calculated as the ratios of the odds of two groups. For example, when examining the effect of a specific map design feature such as \inputvariable{Map Type} (independent variable) on the likelihood of achieving a higher \outputvariable{Source} value (outcome), an odds ratio for \inputvariable{Chloropleth} (Group 1) and \inputvariable{Isarithmic} (Group 2) that is greater than 1 would suggest that the Group 1 odds of yielding a high \outputvariable{Source} value is greater than the Group 2 odds. If the odds ratio is not significantly different from 1, this indicates that there is no difference between the two groups with respect to their impact on the \outputvariable{Source} value.

\begin{figure}[ht]
	\centering
	\includegraphics[width=\linewidth]{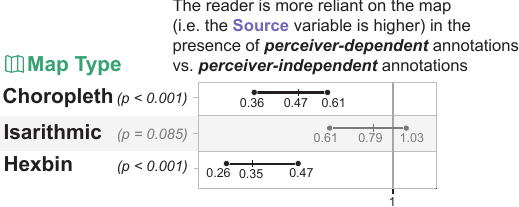}
	\caption{95\% confidence interval chart for the odds ratios model for the \textbf{H1a} hypotheses. The charts summarize the impact of different types of annotations on map reliance in choropleth maps. A notable trend appears: when maps include perceiver-dependent annotations (L3, L4), readers tend to rely less on the map itself, with a likelihood of about 0.47 times compared to maps with perceiver-independent (L2) annotations. In contrast, for maps with perceiver-independent annotations, readers are more than twice as likely ($1/0.47 \approx 2.13$) to rely on the map for their interpretations. This suggests a clear distinction in how different annotations influence reader dependence on the map for understanding the presented data.}
	\label{fig:h1a_95ci}
    \Description{(Bar Plot with Confidence Intervals) The results of the odds ratio statistical test for the H1a hypotheses. When a reader reads choropleth maps or hexbin maps that have perceiver-dependent (PD) annotations versus perceiver-independent (PI), the odds ratios with 95\% confidence intervals were lower than 1. For isarithmic maps with perceiver-dependent (PD) annotations versus perceiver-independent (PI), odds ratio ranged between 0.61 and 1.03. Because this interval spanned a range across 1, this difference was not significant.}
\end{figure}

\subsection{Effects on the Source of Takeaways}

\textbf{Hypothesis 1a (H1a):} \textbf{Map type influences a reader's reliance on text annotations.}

Due to the ordinal nature of the \textbf{\outputvariable{Source}} variable, where higher values indicate more reliance on maps, while lower values show more reliance on text, we used a proportional odds model to analyze the impact of the map and text variables on the comparative odds of being more reliant on map over text (odds ratios). Results from the fitted proportional odds model show that the effect of \inputvariable{Map Type} on \outputvariable{Source} was found to be significantly dependent on the semantic level of text~\footnote{Based on tests on Fixed Effects for main effects and two-factor interaction effects in a proportional odds model with $\alpha=0.05$} \textit{i.e.}, the \inputvariable{Map Type} $\times$ \inputvariable{Semantic Level} interaction effect was significant ($p=0.0004$), but not on \inputvariable{Text-Map Detail Alignment} ($p>0.05$). Estimates of the multiplicative effect on the odds ratios (Fig.~\ref{fig:h1a_95ci})  show that for the choropleth map, perceiver-independent annotations increase the chance of using the map as the primary source of takeaways by a factor of $2.12\times$ ($1/0.47 = 2.12$) when compared to perceiver dependent annotations. This implies that for every 100 people, 32 will report relying more on the map when reading a choropleth map with perceiver-dependent annotations and 68 will report relying more on the map when reading a choropleth map with perceiver-independent annotations. To further clarify, consider a sample of 100 map readers. The proportion of readers that are in the perceiver-dependent group versus perceiver-independent group can be determined by the ratio of 1 to 2.12. When these ratios are normalized (i.e., $\frac{1}{1+2.12}$ for perceiver-dependent and $\frac{2.12}{1+2.12}$) for perceiver-independent annotations), they correspond to approximately 32 and 68 readers. For participants reading an isarithmic map, utilizing perceiver-independent annotations as opposed to perceiver-dependent annotations results in a 1.27$\times$ ($p=0.085$) increase in the odds of predominantly relying on the map. In the scenario of the hexbin map interaction, perceiver-independent annotations result in a 2.86$\times$ ($p<0.001$) increase in the odds of a reader focusing more on the map compared to when perceiver-dependent annotations are present.
Annotations at the dependent level generally increase reliance on text annotations, aligning with the hypothesis's predicted directionality. 

These results indicate that the semantic level's effect (dependent vs. independent) on takeaways is the strongest for isarithmic maps, followed by choropleth maps, and is least apparent for hexbins. This indicates that a higher semantic level tends to foster more reliance on text annotations, with this tendency being most prominent in isarithmic maps and least so in hexbins. 
 
\begin{figure}[t]
	\centering
	\includegraphics[width=\linewidth]{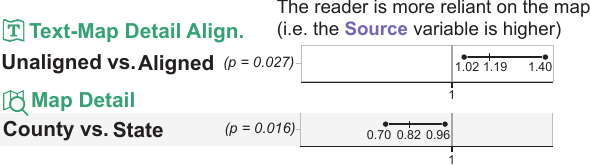}
	\caption{95\% confidence interval chart for the odds ratios model for the \textbf{H1b} hypotheses, which provides an overview of how text alignment and map granularity influence map reliance. This analysis reveals that readers show a $1.19 \times$ higher tendency to rely on maps when text details are unaligned with the map elements, compared to when they are aligned. Furthermore, when comparing state-level choropleth maps with county-level ones, readers show a $1.22 \times$ increased reliance on the state-level maps. These findings highlight the significant role of text alignment and map granularity in shaping how users interact with and interpret map-based information.}
	\label{fig:h1b_95ci}
    \Description{(Bar Plot with Confidence Intervals) The results of the odds ratio statistical test for the H1b hypotheses. For the annotation text-map detail alignment, the one different than the map versus the same as the map had an odds ratio ranging between 1.02 and 1.40, higher than 1. For the map detail, the one with county-level detail versus state-level detail had an odds ratio ranging between 0.7 and 0.96, lower than 1. Both results were significant.}
\end{figure}

\vspace{1.5mm} \noindent \textbf{Hypothesis 1b (H1b):} \textbf{Increased geographic detail in maps leads to greater reliance on text annotations.}

We used a proportional odds model to analyze the impact of the map and text variables on a higher \outputvariable{Source} (\textit{i.e.} reader being more reliant on map). The Fixed Effect tests (Fig.~\ref{tab:fixed-effects}) indicate that \textbf{\inputvariable{Map Detail}} is significant on \outputvariable{Source} ($p=0.0164$). Readers who are shown county-level maps tend to display a higher reliance on text than on the map itself. The odds of a reader relying more on the map for insights are 1.22$\times$ higher ($p=0.016$) when the map detail is at the state level compared to the county level (Fig.~\ref{fig:h1b_95ci}) supporting this hypothesis.

\vspace{1.5mm} \noindent \textbf{Hypothesis 1c (H1c):} \textbf{Higher spatial autocorrelation in maps might reduce reliance on text annotations.}

This hypothesis is rejected. The results reveal that Moran's I is not significant on \outputvariable{Source} for both experiments (Fig.~\ref{tab:fixed-effects}). Similarly, text attributes exhibit no interaction with Moran's I. This leads to the conclusion that differences in geographic correlations within the phenomenon mapped do not influence a reader's reliance on annotations.

\begin{figure}[t]
	\centering
	\includegraphics[width=\linewidth]{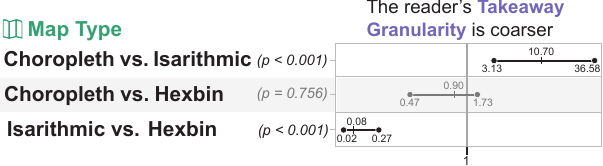}
	\caption{95\% confidence interval chart for the odds ratios model for the \textbf{H2a} hypotheses, which examines the impact of map type on the level of detail in reader takeaways. This chart illustrates a significant trend in how different map types influence the granularity of the information readers extract. When viewing choropleth maps, readers are over 10 times more likely to have a broader, less detailed takeaway compared to isarithmic maps. Similarly, when comparing hexbin maps with choropleth maps, readers lean slightly towards broader takeaways with hexbin maps. The contrast is even more pronounced when hexbin maps are compared with isarithmic maps, where readers are 12.5 times more likely to focus on general, less detailed information. These insights underscore the substantial effect of map type on the level of detail perceived and utilized by readers.}
    \Description{(Bar Plot with Confidence Intervals) The results of the odds ratio statistical test for the H2a hypotheses. The odds ratio of the takeaway granularity is coarser when reading a choropleth map versus reading an isarithmic map ranging between 3.13 and 36.58. The odds ratio of the takeaway granularity is coarser when reading an isarithmic map versus reading a hexbin map ranging between 0.02 and 0.27. The difference was not significant when reading a choropleth map versus a hexbin map.}
    \label{fig:h2a_95ci}
\end{figure}

\subsection{Effects on the Takeaway Granularity}

\textbf{Hypothesis 2a (H2a):} \textbf{Map type affects the granularity of a reader's takeaways.}

The \textbf{\outputvariable{Takeaway Granularity}} output variable, which takes on the values county, state, region is an ordinal data type. We used a proportional odds model to analyze the impact of the input variables on the comparative odds of having a coarser (regional) takeaway. 

We examine the effect of \inputvariable{Map Type} on \outputvariable{Takeaway Granularity}, to which there is a significant effect ($p=0.0003$). According to the odds ratio model which is summarized in Fig.~\ref{fig:h2a_95ci}, readers are 10.70$\times$ ($p<0.001$) more likely to produce takeaways at a coarser granularity when reading a choropleth map vs. an isarithmic map. Readers are also $12.5\times$ more likely to have coarser takeaways when reading a hexbin vs. an isarithmic map. When comparing hexbins to choropleth maps, hexbins were 1.11$\times$ ($p=0.756$) more likely to create coarser takeaways, although this result is not significant. Thus, we conclude that people are least likely to have a coarser-detailed takeaway when reading isarithmic maps.

The immediate implications of these results suggest that choropleth and hexbin maps lead to coarser granularity takeaways, while isarithmic maps help readers focus on finer details. These findings underscore the importance of the contextual use of each map type. For example, in teaching geographic patterns, choropleth maps may be more effective in highlighting trends at the regional level, while isarithmic maps may be better suited for showing phenomena at a finer level.

\begin{figure}[t]
	\centering
	\includegraphics[width=0.9\linewidth]{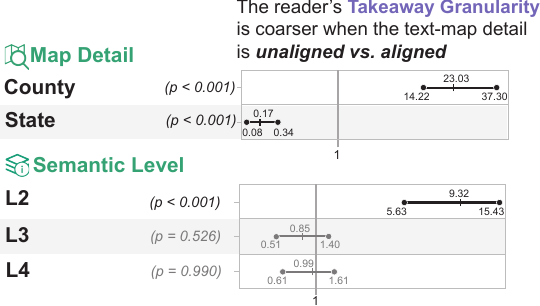}
	\caption{95\% confidence interval chart for the odds ratios model for the \textbf{H2b} hypotheses. All plots model that the reader's takeaway granularity is coarser under different treatments. We show odds ratios for \inputvariable{Map Detail} $\times$ \inputvariable{Text-Map Detail Alignment} and \inputvariable{Semantic Level} $\times$ \inputvariable{Text-Map Detail Alignment}.}
	\label{fig:h2b_95ci}
    \Description{(Bar Plot with Confidence intervals) The results of the odds ratio statistical test for the H2b hypotheses. The Map Detail plot shows the odds ratio of unaligned text-map details versus aligned text-map details ranging between 14.22 and 37.3 when reading a county-level map. This odds ratio became much lower when reading a state-level map, ranging between 0.08 and 0.34. The Semantic Level plot shows only when reading level-2 semantic level annotations, there is a significant relationship between the takeaway granularity and the text-map detail alignment. The odds ratio ranged between 5.63 and 15.43 for the unaligned text-map detail versus the aligned text-map detail when reading level-2 text annotations.}
\end{figure}

\vspace{1.5mm} \noindent  \textbf{Hypothesis 2b (H2b):} \textbf{The granularity of takeaways varies with map detail.}

The test of Fixed Effects (Fig.~\ref{tab:fixed-effects}) suggests that both \inputvariable{Text-Map Detail Alignment} ($p<0.0001$) and \inputvariable{Text-Map Detail Alignment} $\times$ \inputvariable{Semantic Level} ($p<0.0001$) have a significant influence on the granularity of a reader's takeaway. When viewing a county-level map with text-map detail misalignment, the odds of a reader offering a coarser takeaway granularity are 23.03$\times$ ($p<0.001$) greater (Fig.~\ref{fig:h2b_95ci}) than when reading a map with an aligned text-map detail. This result suggests that detailed maps paired with coarser text detail lead to more generalized takeaways. For less detailed (state-level) maps, text detail, regardless of whether it is coarser or finer, tends to elicit more precise (state or county-level) takeaways (Fig.~\ref{fig:h2b_95ci}). It is noteworthy that regardless of the level of text detail, county-level maps yield same-level (county) coarser takeaways (at the state or regional level) compared to state-level maps, since there is no finer level of detail beyond the county level. This suggests that the production of finer-grained takeaways is not caused by higher-detailed maps. 

\begin{figure}[b]
	\centering
	\includegraphics[width=0.5\linewidth]{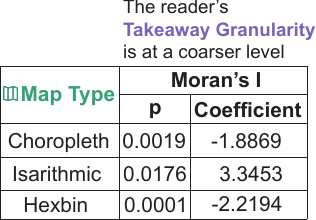}
	\caption{Table of coefficient estimates for the \textbf{H2c} hypotheses. The coefficient shows the correlation between Moran's I and the probability that the \textbf{\outputvariable{Takeaway Granularity}} is coarser. A positive coefficient, as observed with the isarithmic map, indicates a direct correlation between the variables. In the context of an isarithmic map, this means that as Moran's I increases, the reader's takeaway is increasingly likely to be at a coarser level.}
	\label{tab:h2c_table}
    \Description{(Table) The figure shows the table of coefficient estimates for the H2c hypotheses. The table contains Moran's I coefficient values and p-values for the choropleth map, the isarithmic map, and the hexbin map.}
\end{figure}

\vspace{1.5mm} \noindent \textbf{Hypothesis 2c (H2c):} \textbf{Higher spatial autocorrelation leads to takeaways with coarser detail.}

Due to the categorical nature of the dependent variable such as \outputvariable{Takeaway Granularity} and the continuous nature of the independent variable (Moran's I for Spatial Autocorrelation), we use nominal logistic regression to model the odds of having a coarser takeaway. Recall that the ordinal ordering from fine to coarse is county, state, region. 

The effect of Spatial Autocorrelation depends only on \inputvariable{Map Type}, but not on text elements (Fig.~\ref{tab:fixed-effects}). The results indicate that higher levels of spatial autocorrelation produce finer details for both choropleth and hexbin but higher levels of Moran’s I produce coarser level of detail for the isarithmic maps. This is seen by the negative coefficient values for choropleth ($-1.8869$) and hexbin ($-2.2194$) maps, but a positive coefficient ($3.3453$) for the isarithmic map (see Fig.~\ref{tab:h2c_table}). Thus, this hypothesis is only supported for isarithmic maps.

\begin{figure}[t]
	\centering
	\includegraphics[width=\linewidth]{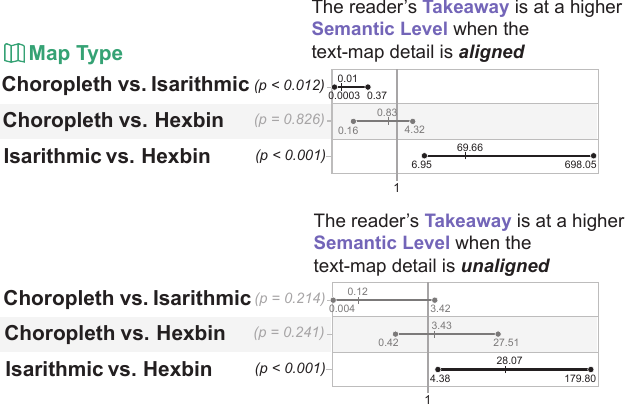}
	\caption{95\% confidence interval chart for the odds ratios model for the \textbf{H3a} hypotheses. The event being modeled is the \textbf{\outputvariable{Takeaway Semantic Level}} having a higher value. Recall that higher semantic levels represent more complex and novel information, while lower levels correspond to basic facts and statistics. The plots show the pairwise comparisons between the three different map types when the text-map detail is aligned (top) and when it is unaligned (bottom).}
	\label{fig:h3a_95ci}
    \Description{(Bar Plot with Confidence intervals) The results of the odds ratio statistical test for the H3a hypotheses. When reading a map with the aligned text details, the choropleth map versus the isarithmic map and the isarithmic map versus the hexbin map had significant results. When reading a map with unaligned text details, only the isarithmic map versus the hexbin map had significant results.}
\end{figure}

\subsection{Effects on the Semantic Level of Takeaways}

\textbf{Hypothesis 3a (H3a):} \textbf{Map type influences the semantic level of a reader’s takeaways.}

To better analyze the semantic level of reader takeaways, which is categorized into ordinal levels (L2, L3, or L4), we again employed a proportional odds model. This model is used to assess the influence of various map and text factors on the likelihood of a reader deriving a takeaway at a more advanced semantic level.

Our model (Fig.~\ref{fig:h3a_95ci}) that estimates the probability that the takeaway will be at a higher semantic level. Fixed Effect tests (Fig.~\ref{tab:fixed-effects}) show that \inputvariable{Map Type} ($p<0.0001$) has an effect on \outputvariable{Takeaway Semantic Level}, but the direction and magnitude of the effect is dependent on the \inputvariable{Text-Map Detail Alignment} (\textit{i.e.} \inputvariable{Text-Map Detail Alignment} $\times$ \inputvariable{Map Type} is also significant at $p=0.0001$). The interpretations of the odds ratios are as follows:

When text and map detail were \textbf{unaligned}, the odds that the reader has a higher semantic level takeaway is 8.33$\times$ ($p=0.214$) when reading an isarithmic map vs. a choropleth map. This difference is more pronounced when text and map detail were \textbf{aligned}, here the odds increased to 100$\times$ ($p=0.012$) for the same comparison - isarithmic map vs. a choropleth map.

When text and map detail were \textbf{unaligned}, the odds that a reader provides a higher semantic level takeaway is 3.43$\times$ ($p=0.241$) greater for a choropleth map vs. a hexbin. When text and map detail were \textbf{aligned}, the odds of readers providing a higher semantic level takeaway is 1.20$\times$ ($p=0.525$) greater for a hexbin vs. a choropleth.

When text and map detail were \textbf{unaligned}, the odds that a reader provides a higher semantic level takeaway is 28.07 $\times$ ($p<0.001$) for an isarithmic map vs. a hexbin. When text and map detail were \textbf{aligned}, the odds increased to 69.66$\times$ ($p<0.001$) for the same comparison - an isarithmic map vs. a hexbin.

These quantitative results show that regardless of the level of text-map detail alignment, isarithmic maps produce significantly higher semantics than choropleths, with the effect stronger when the text detail is the \textbf{aligned}. Thus, this hypothesis is supported. 

\begin{figure}[t]
	\centering
	\includegraphics[width=\linewidth]{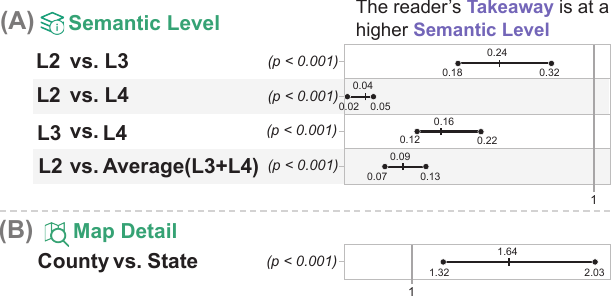}
	\caption{95\% confidence interval chart for the odds ratios model for the \textbf{H3b} hypotheses. The event modeled is that the reader's takeaway will be at a higher semantic level, similar to all other H3 hypotheses. In (A), we compare the odds ratios across different semantic levels. In (B), we examine the odds ratios when comparing map details at the county level vs. the state level, providing insight into how the granularity of map detail influences the semantic level of reader takeaways.}
	\label{fig:h3b_95ci}
    \Description{(Bar Plot with Confidence intervals) The results of the odds ratio statistical test for the H3b hypotheses. The odd ratios for different semantic levels in part A were all below 1. For map detail, the odd ratios ranged between 1.32 and 2.03.}
\end{figure}

\vspace{1.5mm} \noindent  \textbf{Hypothesis 3b (H3b):} \textbf{Coarser map details lead to higher-level semantic takeaways.}
We build an odds model to analyze the impact of map and text factors on the comparative odds of having a higher semantic level takeaway. We see that test of Fixed Effects indicates that \inputvariable{Map Detail} ($p<0.0001$) has a significant effect on the semantic level of the takeaways (Fig.~\ref{tab:fixed-effects}). More specifically, county-level maps produce higher semantic level takeaways over state-level maps by a factor of 1.64$\times$ ($p<0.001$) (Fig.~\ref{fig:h3b_95ci}B). This implies that finer map details are more likely to lead to takeaways at a higher semantic level. The opposite of our hypothesis is true, so it is rejected.

When examining this hypothesis, we also discovered that \inputvariable{Semantic Level} has a highly significant effect on the reader's \outputvariable{Takeaway Semantic Level} ($p<0.0001$). In experiment 2 (Fig.~\ref{fig:h3b_95ci}A), the pairwise comparisons in our odds ratios model yielded: Readers have 4.177$\times$ ($p<0.001$) higher odds of having a higher level semantic takeaway when reading L3 annotations compared to L2 annotations. When reading L4 annotations compared to L2, the odds increase significantly to 25$\times$ ($p<0.001$) higher. Moreover, readers have 6.25$\times$ ($p<0.001$) higher odds of experiencing a higher level semantic takeaway when comparing L4 to L3 annotations. From these findings, we conclude that higher semantics in the text annotations result in higher semantics in the takeaways.  

\begin{figure}[t]
	\centering
	\includegraphics[width=0.55\linewidth]{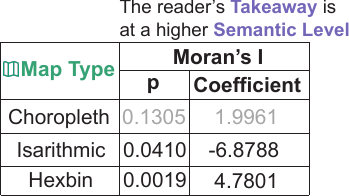}
	\caption{Table of coefficient estimates for the \textbf{H3c} hypotheses. The coefficient shows the correlation between Moran's I and the odds that the \textbf{\outputvariable{Takeaway Semantic Level}} is higher. A positive value indicates that as Moran's I increases, the reader's takeaway will be at a higher semantic level.}
	\label{tab:h3c_table}
    \Description{(Table) The figure shows the table of coefficient estimates for the H3c hypotheses. The table contains Moran's I coefficient values and p-values for the choropleth map, the isarithmic map, and the hexbin map.}
\end{figure}

\vspace{1.5mm} \noindent \textbf{Hypothesis 3c (H3c):} \textbf{Spatial autocorrelation within a map dataset influences the semantic level of takeaways.}

For this hypothesis, we use nominal logistic regression  (Fig.~\ref{tab:h3c_table}) to model how Moran's I affects the semantic level of a reader's takeaway. \inputvariable{Spatial Autocorrelation} $\times$ \inputvariable{Map Type} ($p<0.0001$) has a significant effect on \outputvariable{Takeaway Semantic Level}. For isarithmic maps, higher Moran's I results in lower semantic levels, as denoted by a significant p-value and negative coefficient($coefficient=-6.8788,p=0.0410$). Hexbins produce takeaways at a higher semantic level with higher spatial autocorrelation ($coefficient=4.7801,p=0.0019$). Choropleth maps also are more likely to produce takeaways at a higher semantic level with an increasing Moran's I ($coefficient=1.9961,p=0.1305$), but this result was not significant.

\subsection{Additional Statistical Analyses}
\label{subsection:additional_statistical_analyses}

The preceding subsections focused on examining the influence of map variables on participant takeaways. This subsection aims to concisely present insights gained from analyzing participants' VLAT scores. It's important to note that VLAT scores are observed outcomes; hence, any causal relationships cannot be established based on these scores.

The Pearson correlation coefficient between VLAT scores and the variable \textbf{\outputvariable{Source}} is approximately 0.085 ($p<0.001$), signifying a significant correlation, albeit a slight correlation. This suggests that participants with higher visual literacy are more inclined to extract information directly from the map. However, we reiterate this relationship is correlative and does not imply causation. Furthermore, the nature of the \textbf{\outputvariable{Source}} as ordinal data presents limitations when correlating with VLAT scores, which are cardinal in nature.

Similarly, the Pearson correlation coefficient of -0.033 ($p<0.023$) shows a weak, yet statistically significant negative relationship between VLAT scores and the \textbf{\outputvariable{Takeaway Semantic Level}}. This indicates that higher VLAT scores correlate with lower levels of takeaway semantics. It further suggests that individuals with better visual literacy are inclined to rely less on personal interpretation, favoring direct data extraction (e.g., reading specific values from a map) rather than making abstract or trend-based inferences.

Regarding the volume of takeaways, the mean number recorded per dataset is 774, with a standard deviation of approximately 11.85. This statistic reflects that each dataset in the study, on average, elicited about 774 takeaways. On an individual level, participants contributed an average of 46 takeaways each, with a standard deviation of 12.48.

\section{Discussion}
\label{section:discussion}

\begin{figure*}[t]
\centering	
\includegraphics[width=\linewidth]{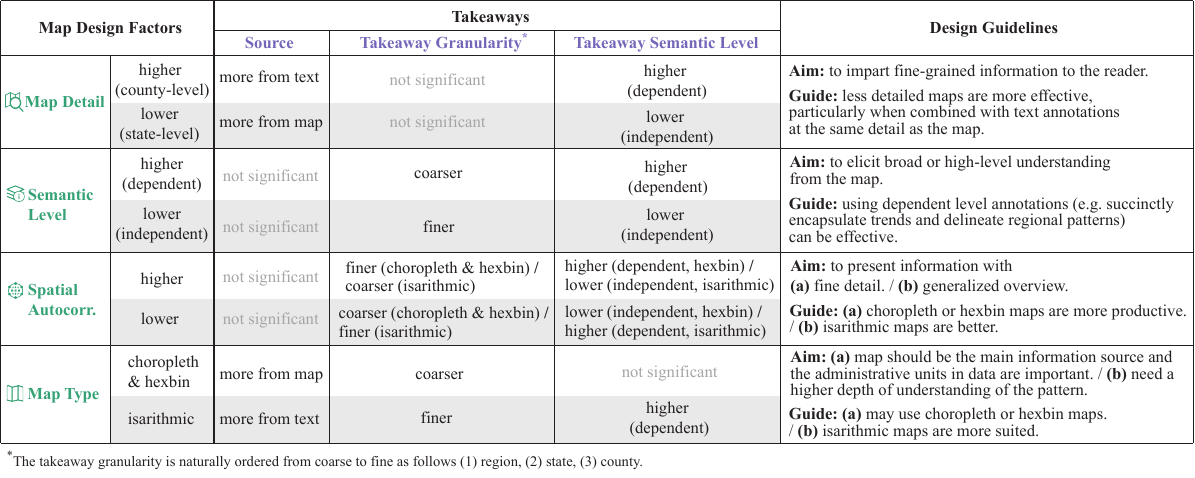}
\caption{Summary of design implications and design guidelines.}
\label{fig:implication_summary}
\Description{(Table) This figure is a table that summarizes the design implications and related design guidelines as Section 5.1 illustrates.}
\end{figure*}

The implications of this study come at an opportune time when digital information sources are more widely used and trusted. Thematic maps, which are useful as statistical reporting tools, are sometimes used to spatially represent critical information in high-stakes domains. Examples include the spread of infectious diseases during a pandemic~\cite{rezaei2020application} and resource distribution in natural disaster mitigation~\cite{FEMAmap2023}. Thus, optimally integrating textual annotations with different thematic maps at varying levels of detail can produce maps that are tailored to the specific readership in focus. The consequences of the findings from this study are explored in this section to provide actionable recommendations for the visualization community. These findings and their associated design implications are summarized in  Fig.~\ref{fig:implication_summary}. It has to be noted that the participants recruited for this study primarily represent a segment of the general population with a background in higher education, with $n=47$ holding 4-year degrees, while some have post-graduate qualifications. They are not experts in cartographic interpretation. Thus, the insights discussed here are intended to aid the comprehension and usability of thematic maps for the educated population.  

\subsection{Design Implications}

\textbf{Effects of Geographic Detail}: As per \textbf{H1b} and \textbf{H3b}, readers (without specialized training in map reading and use) who saw maps with higher geographic detail (\textit{i.e.} county-level maps) tend to have a higher reliance on text and produce higher semantic level takeaways. This could be due to the variation in the expertise of the map reader. \textit{Counterintuitive, our results suggest that is the aim is to impart fine-grained information to the reader, less detailed maps are more effective, particularly when combined with text annotations at the same detail as the map.} A focused approach that emphasizes the regions of interest while potentially reducing the prominence of irrelevant areas can enhance the reader's comprehension. This strategy is often adopted in census maps, where a targeted representation not only prevents information overload but also facilitates a more insightful reading experience.

\textbf{Effects of Semantic Level of Text Annotations}: Based on the outcomes from \textbf{H2b} and \textbf{H3b}, the semantic level of text annotations significantly impacts the granularity and semantic level of takeaways. Annotations with higher semantic levels (dependent) are more likely to produce spatially coarser takeaways and higher semantic level takeaways. Therefore, \textit{when the intention is to elicit broad or high-level understanding, using dependent level annotations can be effective.}
To further contextualize, it is worth noting that annotations that succinctly encapsulate trends and delineate regional patterns stand as powerful tools in encouraging a high-level comprehension. This approach is already employed in various real-world scenarios including, but not limited to, weather forecasting and temperature mapping, where conveying overarching trends and patterns to news viewers takes precedence.

\textbf{Spatial Autocorrelation and Map Type}: Results from \textbf{H1c} and \textbf{H2c} show that spatial autocorrelation does not significantly influence the source of takeaways or reliance on text, but it does influence the detail of takeaways depending on the map type. Higher levels of spatial autocorrelation produce finer details for choropleth and hexbin maps but coarser detail for isarithmic maps. Therefore, \textit{when designing maps, the author should be aware that the type of map chosen will impact the granularity of the takeaways.}
Being aware of the spatial autocorrelation existing in the dataset can guide the optimal choice of map type to communicate details more effectively. For instance, in scenarios where fine detail is important, leveraging the strengths of choropleth or hexbin maps can be more productive. Conversely, if the intention is to present a more generalized overview, isarithmic maps are a better choice.

\textbf{Balancing Visual Complexity with Semantic Detail}: The results from \textbf{H1a} and \textbf{H3a} reveal that map type impacts both the reliance on text and the depth of understanding derived from the map. Isarithmic maps, with their de-emphasis of the administrative units for which data were collected, often lead \address{readers} to seek clarity from text annotations, resulting in a higher level of understanding. Conversely, map types that visually emphasize the administrative units for which data were collected, such as choropleth maps, encourage readers to derive information directly from the map, particularly when paired with more accessible annotations. Regardless of the text-map detail alignment, hexbins and choropleths do not differ significantly with respect to their effect on the semantic level of takeaways. Additionally and as expected, dependent level annotations produce higher level semantics in the takeaways, regardless of the Map Type. 
Therefore, \textit{map creators should strategically choose the map type based on the desired depth of understanding and the reliance on the map itself versus the accompanying text.} Hexbin or choropleth maps are preferable when the map should be the primary source of information, especially in cases where the complexity of information can't be altered. Conversely, isarithmic maps are more suited when a higher depth of understanding is the goal, but they may risk overwhelming the reader.
Given this, it is evident why most online news sources leverage choropleth maps or cartograms — these maps are extensively used in representing census, socioeconomic, or demographic data where delineating administrative units clearly is pivotal. This style accommodates a direct, unambiguous visualization of data corresponding to specific regions. In contrast, isarithmic maps naturally fit scenarios requiring the representation of continuous spatial data distributions, like weather forecasts or temperature maps. This is because they effectively convey gradients and variations, offering a holistic view of data trends across different geographical expanses. Therefore, understanding the inherent strengths of each map type can guide designers in choosing the most effective way to communicate data publicly.

To synthesize, the study illustrates the complex interplay between map design elements and their impact on reader comprehension and information takeaway. Notably, this research highlights the importance of considering the interaction between different map elements, such as type, detail, and spatial autocorrelation, along with the semantic level and detail of accompanying text when designing maps. By strategically utilizing these elements, cartographers can better guide reader comprehension and the granularity and depth of their takeaways.

\subsection{Additional Insights}
\label{subsection:additional_insights}

In this subsection, we refer to participants by their ID from P1-P103.

\vspace{1mm} \noindent \textbf{Subjectivity of Takeaways}: The interpretation of maps can be influenced by personal context and knowledge. For example, P30 writes ``Florida does achieve more employment than Ohio, where I am located. The population might be more homogenous [sic],'' while P47 drew on their knowledge about the U.S. agricultural industry: ``The agricultural sector is highly dependent on government policies and subsidies, which have a significant impact on the prices of agricultural commodities.'' These insights can provide unique perspectives on the data, but also highlight the potential for subjectivity in interpretation.

Some responses provided by participants were not easily verifiable. For example, P2 wrote about the map in Fig.~\ref{fig:teaser}C, ``less vacancy in large metro areas.'' Although this may be true, we surmise that different readers may disagree on the validity of this statement. This example demonstrates how readers may bring their own assumptions to the interpretation of maps, which could lead to unobjective conclusions.

These results indicate that participants bring their own unique perspectives and experiences to the interpretation of visual information. This subjectivity can lead to diverse and creative insights, but can also introduce biases and inaccuracies in interpretation, which can hinder effective decision-making. Therefore, it is important to acknowledge and account for the subjective nature of map reading in research and in practical applications, while also valuing the potential for diverse perspectives and insights.

\vspace{1mm} \noindent \textbf{VLAT Scores}: Participant scores on the VLAT were not considered in the results because the inclusion of these results did not yield any meaningful conclusions. We believed that VLAT scores would correlate positively with the number of takeaways per person. However, it was found that participants would provide the same number of takeaways for each map stimulus throughout the entire duration of the study. Thus, the VLAT section in the study was primarily a mechanism for preventing stimuli fatigue.

\vspace{1mm} \noindent \textbf{Repeating content in annotations}: Our study observed instances where participants echoed the content of annotations in their responses, reflecting the impact of annotations on the interpretation process. This aligns with findings from Hullman et al., highlighting how the presence or absence of annotations can steer viewers' attention and shape their understanding of the visualization~\cite{hullman2011visualization}. However, this effect was small, as only 44 takeaways copied the text annotations from the maps, with 2 takeaways copying Semantic Level 2 annotations, 28 copying Level 3, and 14 copying Level 4. This distribution suggests that the majority of these repetitive takeaways are categorized under semantic levels 3 and 4. The higher occurrence in these levels indicates that while some participants directly mirrored the annotations, they predominantly did so in contexts requiring a more complex understanding (levels 3 and 4), rather than merely restating basic facts (level 2).

\section{Limitations and Future Work}
\label{section:limitations}

\vspace{1mm} \noindent \textbf{Design Factors}: We examined the effects of map type, map detail, spatial autocorrelation, semantic level, and text-map detail alignment on reader takeaways, which is only a subset of thematic map design parameters. Limiting the design options was necessary due to the overwhelming design space of thematic maps. In Section~\ref{subsubsection:thematic_map_design}, we listed other common design factors such as color and classification schemes, both of which were kept constant in our study. Additional design factors such as the underlying data from which isopleth maps are generated (county-level versus state level) and county-level versus state-level hexbins, and the number of annotations can also be considered. Future work can and should explore the effects of these additional factors. 

In this study, we focus solely on hexbin cartograms, a specific variant that represents geographical areas as hexagons. Consequently, our findings may not be generalizable to other cartogram types. 

\vspace{1mm} \noindent \textbf{Study Randomization and Order Effects}: In our study, we implemented a random order of dataset presentation based on a Java randomization function. This approach aimed to minimize potential sequential bias that may be introduced when using a fixed dataset order. However, we recognize that this method may introduce variability due to the lack of control over potential sequence effects. We considered stratified randomization and Latin square designs as alternatives for their ability to distribute dataset types evenly and control for order effects. However, these methods may introduce artificial structuring or reduce the randomness of the conditions, respectively. We chose to show our datasets using a the Java randomized order, but acknowledging that this choice carries increased variability. We suggest that subsequent studies include analyses to examine any order effects and their implications for the study results.

\vspace{1mm} \noindent \textbf{Learning Effects Mitigation:} In both experiments, identical datasets were used. To reduce memory-related biases and learning effects among participants, we systematically altered the input variables, such as map design elements and textual semantic levels, for each experiment. This approach ensured that participants were exposed to distinct maps, albeit derived from the same dataset. For instance, a participant who read a hexbin map featuring perceiver-dependent texts in alignment, would encounter a different map in the second experiment—specifically, a county-level choropleth map with perceiver-independent (L2) annotations. The study's design specifics are illustrated in Fig.~\ref{tab:table-studydesign}. Additionally, to further minimize the impact of participants' recollections, a Visual Literacy Assessment Test (VLAT) was inserted between the two experiments.

\vspace{1mm} \noindent \textbf{Subjectivity in Source Variable}: A potential limitation of our study is the reliance on participants' self-reported assessment to determine the source of their takeaways. This method of subjective evaluation can introduce variability in responses, as individual participants may have differing abilities to accurately recall their cognitive processes in map reading. Future work may address by incorporating more objective measures of source determination, such as eye-tracking or other behavioral indicators that more more accurately assess fixation on map elements.

\section{Conclusion}
\label{section:conclusion}

In an era where public sentiment and response to global and national events are heavily influenced by digital information, it is imperative to establish clear guidelines for visualization tools such as thematic maps. Beyond improving clarity and readability, advancing visualization guidelines is vital for upholding truth and trust in data journalism. This study confirmed how various map configurations with textual annotations affected the quality of reader takeaways. In contrast to previous studies that predominantly examined one variable at a time, our research used a factorial experimental design, which granted insights into more complex effects of both map and textual attributes. Our results therefore provide richer insights into the effect of textual annotations for different map designs, highlighting design synergies or potential antagonisms.

\begin{acks}

This material is based upon work supported by the U.S. Department of Homeland Security under Grant Award Number 17STQAC00001-07-00.
The views and conclusions contained in this document are those of the authors and should not be interpreted as necessarily representing the official policies, either expressed or implied, of the U.S. Department of Homeland Security.

\end{acks}

\bibliographystyle{ACM-Reference-Format}
\bibliography{template}


\end{document}